\documentstyle{elsart}

\input epsf
\input rotate.tex

\begin{document}

\newsavebox{\tabela}

\begin{frontmatter}

\hfill INP 1793/PH
\vspace{3mm}

\title{$\rho$-$\omega$ mixing effects in relativistic heavy-ion collisions}
\thanks{Research supported by 
        the Polish State Committee for
        Scientific Research 2P03B-080-12}

\author{Wojciech Broniowski and Wojciech Florkowski}
\address{H. Niewodnicza\'nski Institute of Nuclear Physics,
         PL-31342 Krak\'ow, Poland}

\begin{abstract}
We show that even moderate excess of neutrons over 
protons in nuclear matter, such as
in ${}^{208}{\rm Pb}$, can lead to large $\rho$-$\omega$ 
mixing at densities of the order of
twice the nuclear saturation density and higher.  
The typical mixing angle is of the order of $10^o$.
% which results in 
The mixing may result in  noticeable shifts of the 
positions and widths of resonances. 
We also analyze temperature effects and find that     
temperatures up to 50~MeV have practically no effect on
the mixing. 
\end{abstract}

\end{frontmatter}

% \addtolength{\baselineskip}{1.0\baselineskip}
Over the past few years considerable theoretical efforts have been made to
understand in-medium properties of mesons \cite{heidelberg,hadrons}. Moreover,
the measurement of dilepton spectra in CERES \cite{ceres} and HELIOS 
\cite{helios} experiments at CERN provides a possibility of verifying various
theoretical predictions concerning light vector mesons \cite{li,cassing}. A number of medium
effects is expected to occur in the $\rho $ and $\omega $ channels: shift of
position of resonances, broadening of their widths, reduction of strength,
emergence of collective branches, or mixing of states which have different
quantum numbers in the vacuum.

The mechanism of mixing induced by the medium has a different origin than
the mixing due to explicit symmetry-breaking terms in the Hamiltonian. It
results from the breaking of the symmetry by the {\em medium}. A well-known
example is the mixing of the scalar-isoscalar $\sigma $ meson with the
longitudinal component of the vector $\omega $ meson \cite{chin}. It is
allowed, since the matter state breaks the Lorentz invariance. Recently
Dutt-Mazumder, Dutta-Roy and Kundu \cite{dutt} analyzed the effects of
isospin asymmetry in matter on $\rho $-$\omega $ mixing. Their analysis,
carried out in the Walecka model \cite{walecka}, implies that at asymmetries
such as in ${}^{208}{\rm Pb}$ and at nuclear saturation 
density the $\rho$ and $\omega$ 
mix with an angle of about $2^o$. 
In this Letter we show that the matter-induced $\rho $-$\omega $ mixing
effects may in fact be much larger. Denote the correlator for vanishing 3-momentum  
\mbox{\boldmath $q$}
in the coupled
neutral $\rho $ and $\omega $ channels by 
\begin{equation}
\Pi^{\alpha \beta }(\nu, \mbox{\boldmath $q$}=0)=\left( 
\begin{array}{cc}
\Pi_\rho ^{\alpha \beta }(\nu) & \Pi_{\rho \omega }^{\alpha \beta}(\nu) \\ 
\Pi_{\rho \omega}^{\alpha \beta}(\nu) & \Pi_\omega ^{\alpha \beta }(\nu)
\end{array}
\right) ,
\label{eq:pi}
\end{equation}
where $\nu$ is the energy
variable. Our approach differs from that of Ref.~\cite{dutt}. Instead of
using a particular model to describe the diagonal parts of (\ref{eq:pi}), we choose a simple
form which can mimic the results of various specific calculations: 
\mbox{$\Pi_v^{\alpha \beta }(\nu)=Z_v^{*-1}
 \left((\nu -i\Gamma_v^{*}/2)^2-m_v^{*2}\right)T^{\alpha \beta}$}
where $v=\rho, \omega$. 
The asterisk denotes the in-medium values of the resonance position, $m_v$,
width, $\Gamma_v$, and the wave-function renormalization, $Z_v$. 
The tensor $T^{\alpha \beta }={\rm diag}(0,1,1,1)$ enters 
for $\mbox{\boldmath $q$}=0$  \cite{chin,jean,shiomi}.  
Our parameterization incorporates 
% in the simplest way the 
basic features of mesons
propagating in nuclear medium, such as the
shift of the resonance position, broadening, and wave-function
renormalization. We find this simple parameterization better for our purpose
of analyzing the nature of $\rho $-$\omega $ mixing. Various specific
calculations
have substantially different predictions for $\Pi_\rho ^{\alpha \beta}$ 
and $\Pi _\omega ^{\alpha \beta }$, depending on physics 
involved \cite{jean,shiomi,brscale,lee,hlqcd,hlsqcd,jinlein,%
leupold,hatsuda1,wolf,frimans,pirner,friman1,rapp,klingl1,klingl2,saito,sttw}.

For the off-diagonal term of the 
correlator, $\Pi _{\rho \omega }^{\alpha \beta }$, responsible for the 
mixing in asymmetric matter, we choose the same mechanism as
in Ref.~\cite{dutt}, namely the interaction of vector mesons
with the Fermi sea of protons and neutrons. This is depicted in 
Fig.~1. The coupling of $\omega $ to the proton and neutron is
equal with the same sign, and the coupling of $\rho^0$ to the proton and
neutron is equal and {\em opposite}.  As a result, the sum of diagrams in
Fig.~1 does not vanish if we have isospin-asymmetric
matter, {\em i.e.} excess of neutrons over 
protons. 
\begin{figure}[tbp]
\vspace{0mm} \epsfxsize = 12.5 cm \centerline{\epsfbox{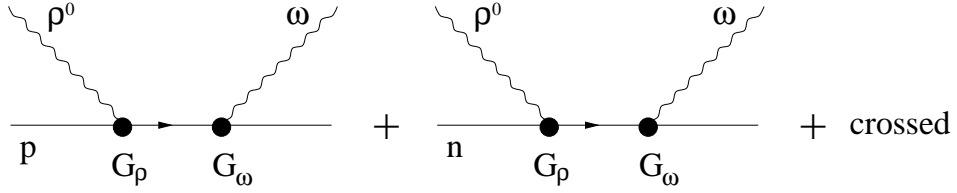}} \vspace{0mm}
%\label{diag}
\caption{Matter-induced mechanism of $\rho$-$\omega$ mixing. 
}
\end{figure}
The coupling of vector mesons to nucleons is described by the Lagrangian
\begin{equation}
L_{{\rm int}}  =   \overline{\psi} \left ( \tau^a G_{\rho, \alpha}
 \rho_a^\alpha \right ) \psi +
 \overline{\psi} \left ( G_{\omega, \alpha} \omega^\alpha \right )
 \psi , \;\;\; 
G_{v, \alpha}  =  g_v \left( \gamma_\alpha -
\frac{\kappa_v}{2M} \sigma _{\alpha \beta } \partial^\beta \right),
\label{lint}
\end{equation}
where $\psi$, $\rho_a^\alpha $ and $\omega^\alpha$ are the nucleon, $\rho $
and $\omega $ fields, and $M$ denotes the nucleon mass in the
vacuum. Note that we include the tensor 
coupling of mesons, which is quantitatively important. 
Following Ref.~\cite{hatsuda1} we use two parameter sets:
\begin{eqnarray}
{\rm I:} & & \;\;\; g_\rho=2.63, \; \kappa_\rho=6.0, \;  g_\omega=10.1, \; 
 \kappa_\omega=0.12, \nonumber \\
{\rm II:} & & \;\;\; g_\rho=2.72, \; \kappa_\rho=3.7, \;  g_\omega=10.1, \; 
 \kappa_\omega=0.12. \nonumber
\end{eqnarray}
This parameterization follows from the vector meson dominance model \cite{vecdom}.
The basic difference between the two sets is the value of $\kappa_\rho$ \cite{kapparho}. 

Applying the usual formalism \cite{chin,hatsuda1} one finds that the 
off-diagonal matrix element in Eq.~(\ref{eq:pi}), describing the mixing 
of $\rho^0$ and $\omega$ (at $\mbox{\boldmath $q$}=0$) is given by
\newpage
\begin{eqnarray}
\Pi_{\rho \omega}^{\alpha \beta}(\nu,\mbox{\boldmath $q$}=0) & = & 
-i \int \frac{d^4k}{(2\pi)^4} \,
\left \{ {\rm Tr}[G_{\rho}^{\alpha}(\nu) S^p_D(k^0+\nu,\mbox{\boldmath $k$})
  G_{\omega}^{\beta}(-\nu) S^p_F(k)] - \right .  \nonumber \\
& & \left . {\rm Tr}[G_{\rho}^{\alpha}(\nu) S^n_D(k^0+\nu,\mbox{\boldmath $k$})
  G_{\omega}^{\beta}(-\nu) S^n_F(k)] \right \}
+ (F \leftrightarrow D) \equiv T^{\alpha \beta} \Pi_{\rho \omega}(\nu) ,
\label{eq:pol}   
\end{eqnarray}
where $S^{p,n}_D(k)$ and $S^{p,n}_F(k)$ denote the density part and the free
part of the Dirac propagator for 
the proton and neutron \cite{chin}. Expression (\ref{eq:pol}) incorporates
the Fermi sea effects and the Pauli blocking. Explicit evaluation gives
\begin{eqnarray}
&& \Pi_{\rho \omega}(\nu) = \frac{2}{3} g_\rho 
g_\omega \int \frac{d^3k}{(2 \pi)^3 E^*(k)} 
\frac{\theta(k_n - \mid \mbox{\boldmath $k$} \mid) -
  \theta(k_p - \mid \mbox{\boldmath $k$} \mid) }
{\nu^2-4E^{*}(k)^2 } \times  \nonumber \\
&& \left [ 8 E^{*}(k)^2+4 M^{*}(k)^2 + 
3 (\kappa_\rho+\kappa_\omega) \frac{M^*}{M} \nu^2 + 
\kappa_\rho \kappa_\omega \frac{E^{*}(k)^2+2 M^{*2}}{M^2} \nu^2 \right ] ,
\label{eq:polexp}   
\end{eqnarray}
where $k_p$ and $k_n$ are the proton and neutron Fermi 
momenta, $M^*$ is the nucleon scalar self-energy in medium,
and \mbox{$E^{*}(k)=\sqrt{M^{*2}+k^2}$}.  
As already mentioned, in symmetric matter, where $k_p = k_n$, the proton 
and neutron contribution to Eqs.~({\ref{eq:pol},\ref{eq:polexp}) cancel and 
 $\Pi_{\rho \omega}(\nu)$ vanishes. In asymmetric matter $k_n > k_p$, and we
get a net contribution to $\Pi_{\rho \omega}(\nu)$. 

The proton and neutron
densities are equal to \mbox{$\rho_{p,n}=k_{p,n}^3/(3 \pi^2)$}, and the 
baryon density $\rho_B$ and the isospin asymmetry $x$ are equal to
\mbox{$\rho_B=\rho_{p}+\rho_{n}$} and \mbox{$x = (\rho_{n}-\rho_{p})/\rho_{B}$}.
At low $x$ it can be easily shown that $\Pi_{\rho \omega}(\nu)$ 
is linear in $x$. 
It remains linear for asymmetries accessible in heavy-ion collisions.
If in addition
we expand Eq.~(\ref{eq:polexp}) at small $\rho_B$, we notice that  
\mbox{$\Pi_{\rho \omega}(\nu) \sim x \rho_B = \rho_n - \rho_p$}, in agreement 
with the low-density theorem for the scattering amplitude \cite{lenz,dover}.

Finding the eigenvalues of the matrix (\ref{eq:pi}) is 
equivalent to solving the following
equation:
\begin{equation}
{\rm Det} \left( 
\begin{array}{cc}
(\nu-i \Gamma_\rho^*/2)^2-m_\rho^{*2} & 
\sqrt{Z_\rho^* Z_\omega^*}\Pi_{\rho \omega}(\nu) \\ 
\sqrt{Z_\rho^* Z_\omega^*}\Pi_{\rho \omega}(\nu) & 
(\nu-i \Gamma_\omega^*/2)^2-m_\omega^{*2}
\end{array} \right) = 0 ,
\label{eq:det}
\end{equation}
where we have moved the wave-function renormalization factors to off-diagonal
terms in order to visualize that the results depend on the product $Z_\rho^*Z_\omega^*$.
Equation (\ref{eq:det}) yields eigenvalues $\nu_1$ and $\nu_2$, 
and the corresponding eigenstates 
$| 1 \rangle$ and $| 2 \rangle$. Our convention is that 
in the absence of mixing, 
{\em i.e.} for $x=0$, we have \mbox{$|1 \rangle = | \rho \rangle$} and 
\mbox{$| 2 \rangle = | \omega \rangle$}.
A commonly used measure of mixing of states is the mixing angle. 
Since the problem (\ref{eq:det}) is not hermitian, the eigenstates
$|1 \rangle$ and $|2 \rangle$
are not orthogonal and we cannot define a single mixing 
angle. We find it useful to 
introduce two mixing angles, $\theta_1$ and $\theta_2$, through the relations
\begin{equation}
| 1 \rangle = \cos \theta_1 | \rho \rangle + 
\sin \theta_1 | \omega \rangle, \;\;\;
| 2 \rangle = -\sin \theta_2 | \rho \rangle + 
\cos \theta_2 | \omega \rangle .
\label{eq:angles}
\end{equation}
Since the matrix in (\ref{eq:det}) is complex, 
the mixing angle are in general complex.

Our results are shown in Table I, which contains 14 
widely different cases. The table should be read 
from top to bottom. The first row labels the case. The second row gives the baryon
density, $\rho_B$, in units of the nuclear saturation density $\rho_0=0.17{\rm fm}^{-3}$. 
Five subsequent rows contain 
our choice of $M^*$, $m^*_\rho$, $m^*_\omega$, $\Gamma^*_\rho$, $\Gamma^*_\omega$ 
and $\sqrt{Z^*_\rho Z^*_\omega}$ for symmetric matter of density $\rho_B$.  
\savebox{\tabela}{\vbox{\begin{tabular}{|l|c|c|c|c|c|c|c|c|c|c|c|c|c|c|}
\hline
 Input:                       & 1 & 2 & 3 & 4 & 5 & 6 & 7 & 8 & 9 & 10 & 11 & 12 & 13 & 14 \\
%\hline
%\multicolumn{15}{|c|}{Input}                       \\
\hline
$\rho_B/\rho_0$               & 2 & 2 & 2  & 2 & 2  & 2 & 2 & 2 & 2 & 2 & 3 & 3 & 2 & 2  \\
$M^*/M$                       &0.5&0.5&0.5 &0.5&0.5 &0.5&0.5&0.5&0.5&0.5&0.4&0.4&0.5&0.5 \\
$m^*_\rho$ (MeV)                   &500&550&650 &500&500 &500&500&550&550&650&550&550&550&550 \\
$m^*_\omega$ (MeV)                 &500&450&450 &500&500 &500&500&450&450&450&450&450&450&450 \\
$\Gamma^*_\rho$ (MeV)               & 0 & 0 & 0  &200&200 &300&200&300&300&300&400&400&300&300 \\
$\Gamma^*_\omega$ (MeV)             & 0 & 0 & 0  & 50& 50 & 50&200& 50&200&200&300&300& 50&200 \\
$\sqrt{Z^*_\rho Z^*_\omega}$  &0.7&0.7&0.35&0.7&0.35&0.7&0.7&0.7&0.7&0.7&0.5&0.3&0.7&0.7 \\
%\hline
%\multicolumn{15}{|c|}{Output for $x=x_{Pb}$}       \\
\hline 
\multicolumn{1}{|c|}{Output, $x=x_{Pb}$:} & \multicolumn{12}{|c|}{Set I} 
 & \multicolumn{2}{|c|}{Set II} \\ 
\hline
${\rm Re}(\nu_1)$  (MeV)           &535&562&652 &509&502 &505&532&557&559&656&559&553&555&557 \\
${\rm Re}(\nu_2)$ (MeV)            &469&442&449 &494&499 &497&470&445&443&446&442&447&446&444 \\
$2{\rm Im}(\nu_1)$ (MeV)            & 0 & 0 & 0  &170&193 &286&218&298&306&309&414&405&297&302 \\
$2{\rm Im}(\nu_2)$ (MeV)            & 0 & 0 & 0  & 83& 57 & 67&186& 56&198&198&294&298& 55&200 \\
${\rm Re}(\theta_1)$ (deg)         &45 &19 & 6  & 12&  3 &  5& 45& 11& 17& 11& 18&11  &  9& 14 \\
${\rm Re}(\theta_2)$  (deg)         &45 &15 & 4  & 11&  2 &  4& 45&  8& 13&  7& 14&9  &  7& 12 \\
${\rm Im}(\theta_1)$  (deg)         & 0 & 0 & 0  &-31&-13 &-15&  0& -7& -1&  2&  1&1  & -6& -2 \\
${\rm Im}(\theta_2)$  (deg)         & 0 & 0 & 0  &-32&-13 &-16&  0& -6& -2&  0&  0&0  & -6& -2 \\
\hline
\end{tabular}
}}
\begin{table}[t]
\vspace{4.5cm}
%\begin{center}
\rotl{\tabela}
%\usebox{\tabela}
%\end{center}
\caption{$\rho$-$\omega$ mixing in asymmetric matter. See text for details.}
\label{tab:results}
\end{table}
As already mentioned, the values for these quantities do not follow from any specific  
calculation, but are chosen according to experience accumulated by various 
existing calculations in the literature \cite{jean,shiomi,brscale,lee,hlqcd,hlsqcd,jinlein,%
leupold,hatsuda1,wolf,frimans,pirner,friman1,rapp,klingl1,klingl2,saito,sttw}. 
Most of our cases take $\rho_B=2\rho_0$. We assume that at this density the nucleon mass drops to 50\% 
of its vacuum value. 
The values of $m^*_\rho$ and $m^*_\omega$ are assumed to be between 450MeV and 650MeV. The widths
take very different values.
The wave function renormalization factors are assumed to be $\sqrt{Z^*_\rho Z^*_\omega}=0.7$, or lower.
Model calculations show that these factors get reduced rather significantly in nuclear medium. 
It is assumed that the coupling constants $g_v$ and $\kappa_v$ do
not depend on density.   
The seven bottom rows of the table 
contain our results for the mixing at asymmetry $x$ such as in ${}^{208}{\rm Pb}$, {\em i.e.}
\mbox{$x=x_{Pb}\equiv 44/208$}, and for parameter sets I and II.
The quantities $\nu_1$ and $\nu_2$ are the complex positions of the mixed states (cf. explanations
after Eq.~(\ref{eq:det})). The value of ${\rm Re}(\nu_1)$ should be compared to $m^*_\rho$, 
${\rm Re}(\nu_2)$ to $m^*_\omega$, $2{\rm Im}(\nu_1)$ to $\Gamma^*_\rho$ and $2{\rm Im}(\nu_2)$
to $\Gamma^*_\omega$. The mixing angles are defined in Eq.~(\ref{eq:angles})

The first three cases assume that the $\rho$ and $\omega$ mesons have zero width, hence are of academic
rather than practical interest. We list them, however, since the assumption of of no width 
has been made in a number of calculations, such as the QCD sum rule calculations 
\cite{hlqcd,hlsqcd,jinlein}, or in the Walecka-model \cite{jean,shiomi,hatsuda1}. Case~1 assumes equal 
vector meson masses in symmetric matter, $m^*_\rho=m^*_\omega=500{\rm MeV}$. In this 
case in asymmetric medium we have ideal mixing, with $\theta_1=\theta_2=45^o$, which results
in splitting of masses by 66MeV. If $m^*_\rho$ and $m^*_\omega$ are split by $100{\rm MeV}$
(case~2), then the mixing angles are equal to $19^o$ and $15^o$, and additional splitting
due to mixing is equal to 20MeV, a 20\% 
effect compared to $m^*_\rho-m^*_\omega$. The increase of $m^*_\rho-m^*_\omega$ to 200 MeV and
simultaneous reduction of the wave-function renormalization (case 3) causes the reduction
of the mixing angles to $6^o$ and $4^o$. Note that such angles are still 
an order of magnitude larger than the mixing angle resulting from the electro-magnetic mixing.
The remaining case have finite width of vector mesons. Cases~4-6 have $m^*_\rho=m^*_\omega$, 
large $\Gamma^*_\rho$ and narrow $\Gamma^*_\omega$. We note that $2{\rm Im}(\nu_2)$ is 
significantly increased compared to $\Gamma^*_\omega$ in cases~4 and 6, 
where $\sqrt{Z^*_\rho Z^*_\omega}=0.7$. The mixing angles have large imaginary parts. 
Case~7 has $m^*_\rho=m^*_\omega$ and $\Gamma^*_\rho=\Gamma^*_\omega=200{\rm MeV}$. In this
case the mixing is ideal, with the mixing angle at $45^o$. Resonance positions and widths are
shifted significantly. Cases~8-9 have $m^*_\rho-m^*_\omega=100{\rm MeV}$, wide $\rho$ and 
narrow (case~8) or wide (case~9) $\omega$. Case~10 has $m^*_\rho-m^*_\omega=200{\rm MeV}$. 
Cases~11-12 take $\rho_B=3\rho_0$, and, correspondingly, lower $M^*$, wider mesons,  
and smaller $\sqrt{Z^*_\rho Z^*_\omega}$. In cases~8-12 the mixing angles are in the range $10-20^o$.
Cases~13-14 are for parameter set II, and should be directly compared to cases~8-9. We can see
that the choice of parameters has very little influence on the mixing effect.  
To summarize, the mixing effects are sizeable for all sensible cases, with mixing angles 
of the order of $10^o$, or larger.

In order to be relevant for the physics of heavy-ion collisions, the above 
analysis has to be extended to finite temperatures.
The most important effect of finite temperature is the production of pions.
We model these effects by assuming that we have ideal gas of protons, neutrons and pions
in thermal equilibrium. The equilibrium with respect to the 
reactions \mbox{$p \leftrightarrow n + \pi^+$}
and \mbox{$n \leftrightarrow p + \pi^-$} 
leads to the following relations among the chemical potentials
for various components of the 
system: \mbox{$\mu_p=\mu_n+\mu_{\pi^+}$}, \mbox{$\mu_n=\mu_p+\mu_{\pi^-}$}.   
The densities of particles are therefore given by
\begin{equation}
\rho_{p,n}(T)  =  2 \int \frac{d^3k}{(2 \pi)^3} 
\frac{1}{e^{\frac{\sqrt{M^{*2}+k^2}-\mu_{p,n}}{T}}+1} , \;\;\;
\rho_{\pi^\pm}(T)  =  \int \frac{d^3k}{(2 \pi)^3} 
\frac{1}{e^{\frac{\sqrt{m_\pi^{2}+k^2} \mp (\mu_p - \mu_n)}{T}}-1}.
\label{eq:tdis}
\end{equation}
There are two constraints in the system: the value of the 
baryon density and the electric charge density, which 
gets the contribution from protons and charged pions. Hence
\begin{equation}
\rho_p(T)+\rho_n(T)  =  \rho_B, \;\;\;
\rho_p(T)+\rho_{\pi^+}(T)-\rho_{\pi^-}(T)  =  \rho_p(T=0) \equiv \frac{1-x}{2} \rho_B, 
\label{eq:finT}
\end{equation}
where we have used the definition of $x$ in the last equality. 
We solve Eq.~(\ref{eq:finT}) numerically
for $\mu_p$ and $\mu_n$ at fixed $\rho_B$ and $x$. 
Equipped with the chemical potentials we can now calculate $\Pi_{\rho \omega}(\nu)$ at finite $T$. 
This is done by applying formula (\ref{eq:polexp}) with the replacement \cite{kapusta,rehberg}
\begin{equation}
\theta(k_{p,n} - \mid \mbox{\boldmath $k$} \mid) \rightarrow
( {e^{\frac{\sqrt{M^{*2}+k^2}-\mu_{p,n}}{T}}+1} )^{-1} .
\label{eq:repl}
\end{equation}
The temperature effects can be described as follows. As we increase $T$,
we produce pions, with the excess of $\pi^-$ over $\pi^+$. The effect starts 
to be significant around \mbox{$T\sim m_\pi\sim 140{\rm MeV}$}. The charge conservation 
constraint causes the reduction of the excess of neutrons over protons. 
As a result, the mixing element $\Pi_{\rho \omega}(\nu)$ is reduced.
Quantitatively, there is practically no effect for $T$ up to 50MeV. At $T\sim 140{\rm MeV}$
 $\Pi_{\rho \omega}(\nu)$ is reduced by about a factor of 2. This occurs for 
a variety of values of $\nu$, $M^*$ and $\rho_B$.
Our analysis shows that the mixing effect will continue to be important at moderate temperatures. 
At higher temperatures, close to the chiral transition, $\Pi_{\rho \omega}(\nu)$ is 
strongly suppressed, but our 
approach is no longer valid in that domain. Note that at higher $T$ also 
the values of $m^*_v$, $\Gamma^*_v$, and $Z^*_v$ are strongly modified.  

Clearly, other processes than that of Fig.~1 can also contribute to in-medium
$\rho$-$\omega$ mixing, {\em e.g.} processes involving pions. 
It is unlikely, however, that such processes cancel the mechanism analyzed here.
Thus, we expect that our results of large $\rho$-$\omega$ mixing 
will show up, among other possible medium-induced effects, in future high-accuracy
relativistic heavy-ion experiments planned at GSI and CERN.

% \bibliography{mix}
% \bibliographystyle{npsty}

\end{document}